\documentclass{aastex6}
\usepackage{amsmath}

\begin{document}
\title{The KBC Void:  Consistency with Supernovae Type Ia and the Kinematic SZ Effect in a $\Lambda$LTB Model}

\author{
Benjamin~L.~Hoscheit\altaffilmark{1,2} and 
Amy~J.~Barger\altaffilmark{2,3,4}
}

\affil{$^1$Department of Physics, University of Wisconsin-Madison, 1150 University Avenue, Madison, WI 53706, USA\\
$^2$Department of Astronomy, University of Wisconsin-Madison, 475 N. Charter St., Madison, WI 53706, USA\\
$^3$Department of Physics and Astronomy, University of Hawaii, 2505 Correa Road, Honolulu, HI 96822, USA\\
$^4$Institute for Astronomy, University of Hawaii, 2680 Woodlawn Drive, Honolulu, HI 96822, USA}

\slugcomment{Accepted to The Astrophysical Journal}

\begin{abstract}
There is substantial and growing observational evidence from the normalized luminosity density in the near-infrared that the local universe is underdense on scales of several hundred megaparsecs.  We test whether our parameterization of the observational data of such a ``void" is compatible with the latest supernovae type Ia data and with constraints from line-of-sight peculiar-velocity motions of galaxy clusters with respect to the cosmic microwave background rest-frame, known as the linear kinematic Sunyaev-Zel'dovich (kSZ) effect. Our study is based on the large local void (LLV) radial profile observed by Keenan, Barger, and Cowie (KBC) and a theoretical void description based on the Lema\^{i}tre-Tolman-Bondi model with a nonzero cosmological constant ($\Lambda$LTB). We find consistency with the measured luminosity distance-redshift relation on radial scales relevant to the KBC LLV through a comparison with 217 low-redshift supernovae type Ia over the redshift range $0.0233 < z < 0.15$.  We assess the implications of the KBC LLV in light of the tension between ``local" and ``cosmic" measurements of the Hubble constant, $H_{0}$. We find that when the existence of the KBC LLV is fully accounted for, this tension is reduced from $3.4\sigma$ to $2.75\sigma$.  We find that previous linear kSZ constraints, as well as new ones from the South Pole Telescope and the Atacama Cosmology Telescope, are fully compatible with the existence of the KBC LLV.  
\end{abstract}

\keywords{cosmology: miscellaneous - methods: numerical - supernovae: general}

\section{Introduction}
\label{secIntro}

Keenan, Barger, \& Cowie (2013; hereafter KBC) have reported evidence that the local universe is underdense on scales extending to $\sim300$~Mpc.  KBC made detailed measurements of the $K$-band luminosity density as a function of distance over radial distances $\sim50-1000$~Mpc using intermediate to high-redshift datasets from the UKIDSS Large Area Survey, the Two-degree Field Galaxy Redshift Survey (2dF-GRS), the Galaxy And Mass Assembly Survey (GAMA), and other surveys. They complemented these with a low-redshift 
dataset from the 2M++ catalog. KBC concluded that out to $z\sim$0.07, there is evidence for a rising luminosity density that, by $z>0.1$, reaches a value $\sim$1.5 times higher than that measured locally.  Similarly, Whitbourn \& Shanks (2014) studied galaxy redshift data  from the Six-degree Field Galaxy Redshift Survey (6dFGS), the Sloan Digital Sky Survey (SDSS), and GAMA, in addition to sky-averaged peculiar velocities of the same areas, and concluded that the Local Group inhabits a ``Local Hole" in the large-scale structure (LSS) of the universe.  To differentiate from the Local Void that is known to lie adjacent to the Local Group (Tully \& Fisher 1987), we introduce the terminology small void ($<100$~Mpc), large void (100--1000 Mpc), and giant void ($>$~Gpc scale). The void found by KBC is then a large local void (LLV).  A recent study by Chiang et al. (2017) using standard candles found supporting evidence for the KBC local inhomogeneity.  

With such substantial and growing observational evidence for an LLV, there has been increased interest in probing cosmological 
dynamics on a local scale, namely, through measurements of the local expansion rate of the Universe, $H_{0}$.  
The recent 2.4\% local measurement using distance-ladder-based techniques (i.e., through the megamaser-determined distance to NGC4258, Cepheid variable star period-luminosity measurements, and the supernovae type Ia (hereafter, SNe Ia) luminosity distance-redshift relations) is $H_{0}$ = 73.24~$\pm$ 1.74 km {$\rm s^{-1}$~$\rm Mpc^{-1}$} (Riess et al.\ 2016; hereafter, R16).  However, the cosmic measurement from the \textit{Planck} satellite data is $H_{0}$ = 66.93$\pm$ 0.62~km {$\rm s^{-1}$~$\rm Mpc^{-1}$} (Ade et al.\ 2014).  
The discrepancy between the local and cosmic values is 3.4$\sigma$; this is commonly referred to as the Hubble constant tension.  

R16 explored whether systematic contributions from inhomogeneity in the local LSS on redshift scales relevant to the low-redshift SNe Ia measurements could potentially be causing the tension in $H_{0}$.  In particular, R16 presented a $\Delta H/ H$ plot averaged over a fine sequence of redshift bins of size $\Delta z = 0.15$ in the redshift range $0.0233 <z< 0.4$. R16 dismissed the possibility of an LLV, as they did not find evidence for a systematic change in $H_{0}$ with redshift bin.  However, we present evidence below that a binning of their data internal ($0.0233 <z< 0.07$) and external ($0.07 <z< 0.15$) to the KBC LLV shows evidence for a drop in $H_{0}$ of 
1.27 $\pm$ 0.59 km {$\rm s^{-1}~\rm Mpc^{-1}$}, which we interpret as a consequence of an LLV model.  In turn, this softens the discrepancy between the $H_{0}$ external to the void and the \textit{Planck} $H_{0}$ to a significance of $2.75\sigma$.

A simplified model to illustrate the effect of an LLV on the Hubble constant is to take a top-hat distribution for the void mass density deficit. The expression for the luminosity distance, $D_{L}$, internal and external to the LLV is then
\begin{equation} 
 \frac{D_{L}(z)}{1 + z} =
  D_{H} \begin{cases} 
      \hfill \frac{1}{\sqrt{{\Omega_{k, in}}}} {\rm sinh}[ \sqrt{{\Omega_{k, in}}} \int_{0}^{z} \frac{dz'}{\sqrt{\Omega_{M, in} (1 + z')^{3} + \Omega_{k, in} (1 + z')^{2} +  \Omega_{\Lambda, in}}}]    \hfill & \text{ z $\leq$ $z_{edge}$} \\
      \hfill  \int_{0}^{z} \frac{dz'}{\sqrt{\Omega_{M, out} (1 + z')^{3} +  \Omega_{\Lambda, out}}}  \hfill & \text{ z $>$ $z_{edge}$} \\
  \end{cases} \,,
  \label{simpledL}
\end{equation}
where we have made the distinction between cosmological parameters internal and external to the LLV with the labels ``in" and ``out".  
We take $z_{edge} = 0.07$ and $D_{H}$ = ${c}/{H_{0}}$ = 4096 $\pm$ 97.2 $\rm Mpc$.   

External to the LLV, we assume the energy density parameters are consistent with the best fitting 
\textit{Planck} parameters in the FRW cosmology with cold dark matter and a 
cosmological constant $\Lambda$ ($\Lambda$CDM);
i.e., $\Omega_{M, out}$ = 0.3, $\Omega_{\Lambda, out}$ = 0.7, and $\Omega_{k, out} = 0$ (Ade et al.\ 2014). 
Internal to the LLV, we take $\Omega_{M, in} = 0.21$ to describe the KBC LLV.  Since the sum of the three energy density 
parameters is constrained to be one, we have only one free energy density parameter internally, either $\Omega_{k, in}$ or 
$\Omega_{\Lambda, in}$.

\begin{figure*}
\center{\includegraphics[width=13cm,angle=0]{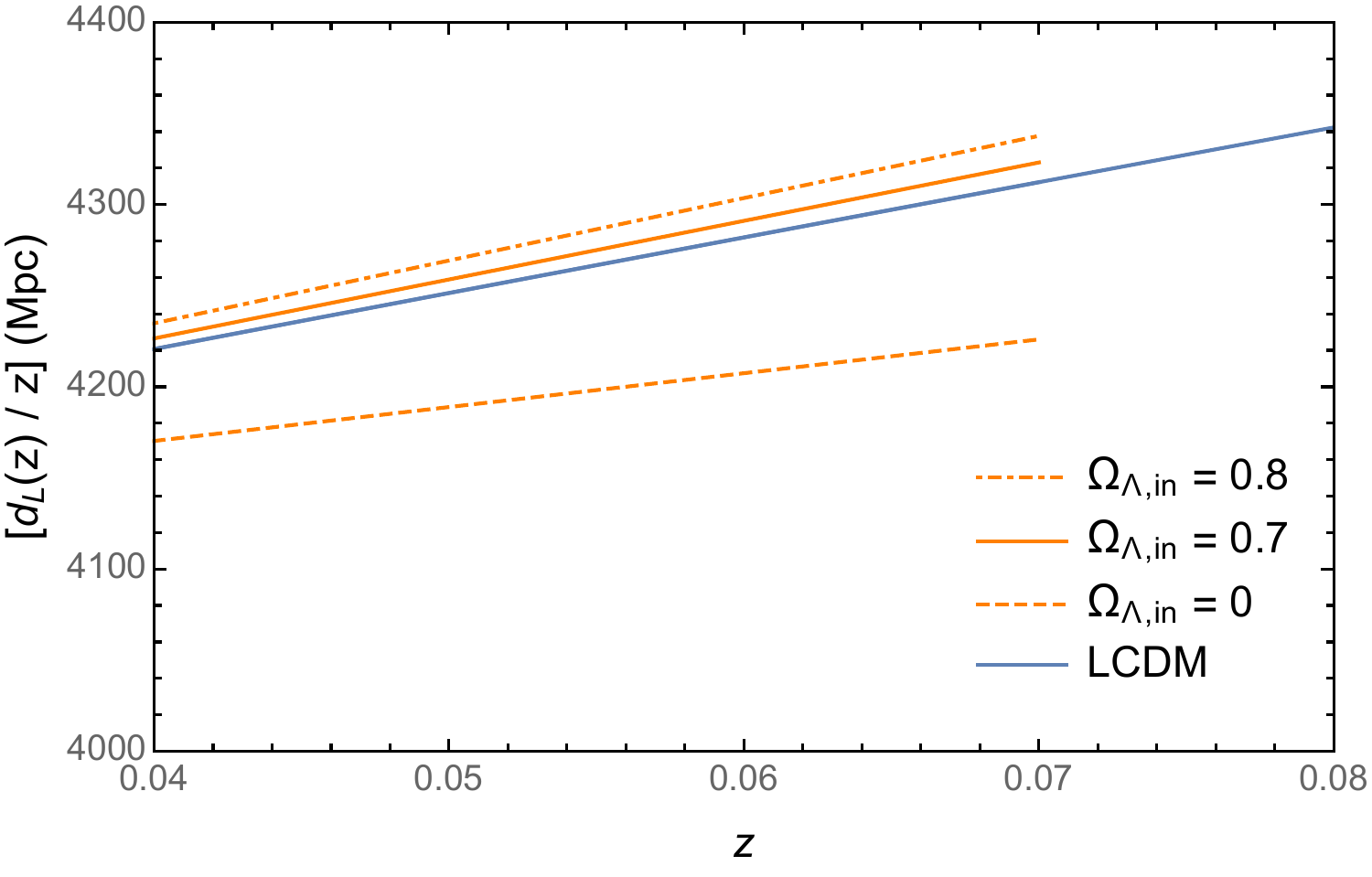}}
\caption{Luminosity distance divided by redshift vs. redshift for the simple toy model LLV of 
Equation~\ref{simpledL} 
with $z_{edge} = 0.07$, $\Omega_{M, in}=0.21$,
several values of $\Omega_{\Lambda, in}$ (orange lines; see legend), and
$\Omega_{k, in}=1-\Omega_{\Lambda, in}-\Omega_{M, in}$.
We use $H_{0}$ = 73.2 km {$\rm s^{-1}$ $\rm Mpc^{-1}$}.
External to the LLV, the luminosity distance is described by the concordance 
cosmological model (blue line) with $\Omega_{M, out}=0.3$ and
$\Omega_{\Lambda, out}= 0.7$; here, we have extrapolated it into the void.
}
\label{simpledLpic}
\vskip 0.5cm
\end{figure*}

In Figure~\ref{simpledLpic}, we show the luminosity distance from Equation~\ref{simpledL}
divided by redshift versus redshift for three choices of $\Omega_{\Lambda, in}$ (orange lines). 
We also show the luminosity distance described by $\Lambda$CDM (blue line), 
which applies external to the LLV; here, we have extrapolated it into the void.
Including the KBC LLV does not cause a large change relative to $\Lambda$CDM
across the void boundary, even for the extreme assumption of $\Omega_{\Lambda, in} =0$.  
For the choice $\Omega_{\Lambda, in} =0.7$, the KBC LLV produces a change in the distance 
modulus ($\mu = \text{m} - \text{M}$) of only 0.005, while for $\Omega_{\Lambda, in} =0$, 
it produces a change of $-0.04$.

The structure of the paper is as follows.
In Section~\ref{secKBC}, we model the dynamics of the KBC LLV using a radially inhomogeneous Lema\^{i}tre-Tolman-Bondi spacetime (e.g., Sundell et al.\ 2015) with matter and a cosmological constant (hereafter, $\Lambda$LTB) that eventually converges back to a standard spatially flat Friedmann-Robertson-Walker spacetime with matter and a cosmological constant (hereafter, $\Lambda$FRW). In Section~\ref{secSNe}, we show that the KBC LLV is consistent with the low-redshift SNe Ia data and contributes to the resolution of
the Hubble constant tension.
In Section~\ref{seckSZ}, we examine the theoretical framework for the linear kinematic Sunyaev-Zel'dovich (kSZ) effect, which results from line-of-sight peculiar-velocity motions of galaxy clusters with respect to the cosmic microwave background (CMB) rest frame (Sunyaev et al.\ 1980). Local void models have been considered as a potential explanation of the apparent cosmic acceleration inferred from SNe Ia data.  
However, several groups have used constraints from the detection of secondary CMB anisotropies to rule out a class of giant (Gpc-scale) local void (GLV) models (e.g., Garc\'{i}a-Bellido et al.\ 2008; Zhang et al.\ 2011, 2015; Moss et al.\ 2011; Zibin \& Moss 2011; Bull et al.\ 2012), since the GLV models that were able to fit the SNe Ia distance modulus data predicted far more kSZ power than allowed by the 95\% confidence upper limits posted by the South Pole Telescope (SPT) and the Atacama Cosmology Telescope (ACT) (Hall et al.\ 2010; Das et al.\ 2011).  In Section~\ref{secConstraints}, we show that the KBC void, which is an LLV instead of a GLV, is consistent with previously used and new linear kSZ constraints. 
In Section~\ref{secSummary}, we summarize our results.

\section{Parameterizing the KBC LLV}
\label{secKBC}
In this section, we parameterize the KBC LLV and introduce the LTB inhomogeneous cosmological formalism. We assume spherical symmetry and an observer located at the symmetric center of the underlying mass density distribution of the KBC LLV.  A rough positional coincidence of the void's center with the observer's location is presumed to be accidental. Note that we are averaging over inhomogeneities in the distribution of matter (i.e., galaxies and larger scale structure) within the LLV.  Relaxing the assumption of sphericity is beyond the scope of our present study.  For instance, one may relax spherical symmetry in a quasispherical Szekeres model by allowing the observer to be off-center in an otherwise spherically symmetric void (Sussman et al.\ 2012; Bolejko et al.\ 2016).  

Following KBC, we assume that measurements of the $K$-band luminosity density accurately probe the underlying mass density distribution locally (Marinoni et al.\ 2005; Maller et al.\ 2005).  More precisely, we take $b$ = 1, where $b$ is the linear bias parameter. As noted in KBC, Maller et al.\ (2005) measures $b_{K}(z\approx 0) = 1.1 \pm 0.2$. We convert the $K$-band luminosity densities to density contrast values using the conversion scale established in KBC based on Bolejko et al.\ (2011). Then, adopting $H_{0}$ = 73.2~km~{$\rm s^{-1}$~$\rm Mpc^{-1}$} from R16, we can plot the density contrast versus comoving distance. 

\begin{figure*}
\center{\includegraphics[width=13cm,angle=0]{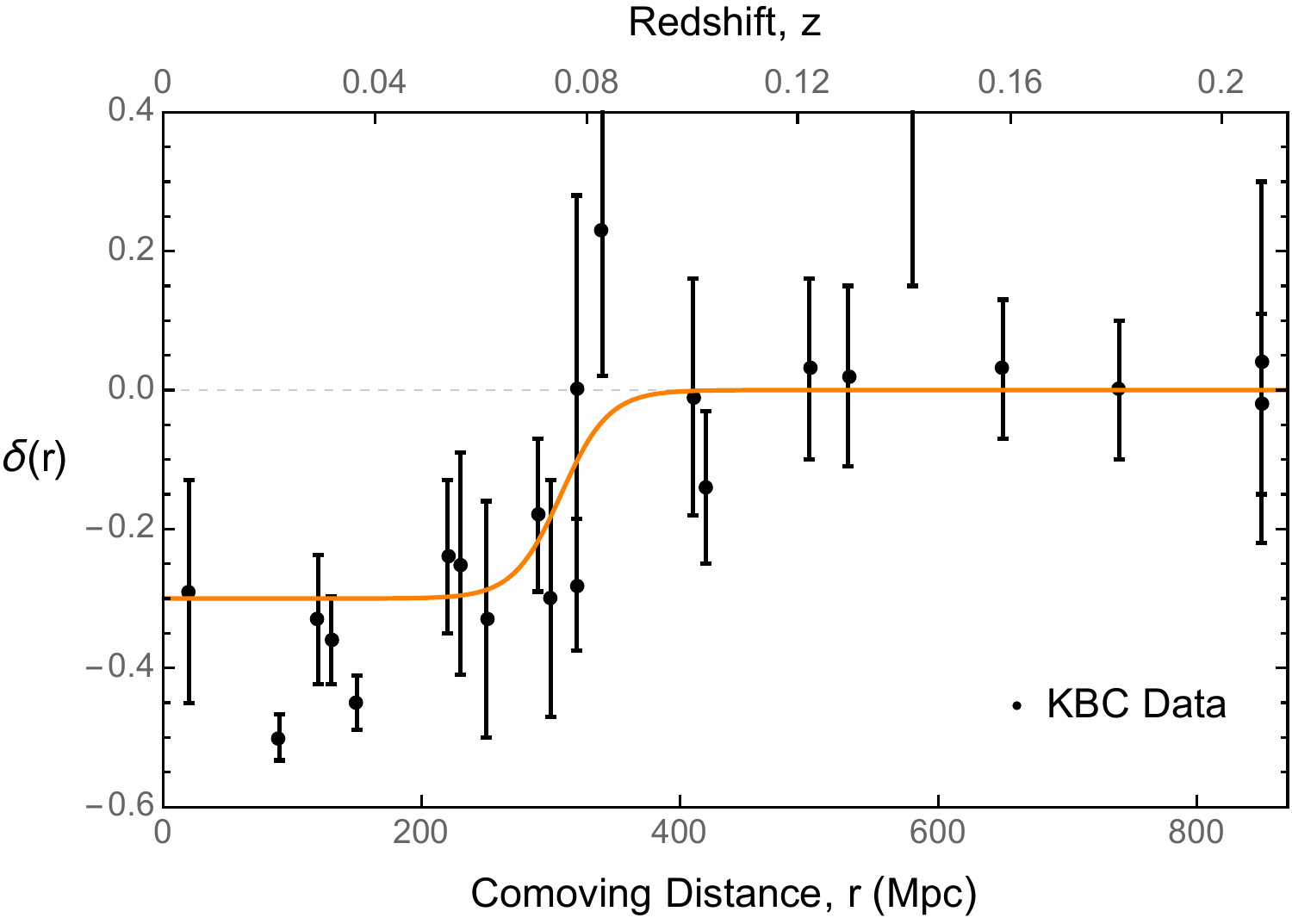}}
\caption{A CGBH mass density parameterization (orange curve; $\delta_{V}$ = $-0.3$, $r_{V} = 308$~Mpc, and $\Delta r = 18.46$~Mpc) of the density contrast data from KBC (black points).  This is a representative description of the void mass density profile and not a $\chi^{2}$-fit to the data points.  There is a radial increase in density contrast around 250$-$350~Mpc, beyond which the mass density normalizes to that of a homogeneous background.  We adopt $H_{0}$ = 73.2 km {$\rm s^{-1}$ $\rm Mpc^{-1}$} and assume a linear Hubble law conversion from comoving distance to redshift.}
\label{figData}
\vskip 0.5cm
\end{figure*}

In Figure~\ref{figData}, we show that the observed increase in the radial mass density distribution from KBC (black points) is reasonably well described by a constrained Garc\'{i}a-Bellido Haugb$\o$lle (CGBH) parameterization (Garc\'{i}a-Bellido et al.\ 2008b) (orange curve) given by  
\begin{equation} 
\delta(r) = \delta\textsubscript{V}\Biggl(\frac{1 - {\rm tanh}[(r - r\textsubscript{V})/2\Delta r]}{1 + {\rm tanh}(r\textsubscript{V}/2\Delta r)}\Biggr) \,.
\end{equation} 
In terms of the spherically averaged mass density within the void, $\overline{\rho}(r)$, and the the mass density at the homogeneity scale, $\rho\textsubscript{0}$, the fractional deficit is given by
\begin{equation}
\delta(r) \equiv (\overline{\rho} (r) - \rho\textsubscript{0}) / \rho\textsubscript{0} \,.
\end{equation}
In Equation 2, $\delta_{V}$ describes the mass density at the symmetric center, $r_{V}$ is the characteristic size of the void, and $\Delta r$ describes the steepness of the void near the edge (Zhang et al.\ 2015).  
Our $K$-band data are not yet sufficient for a statistical analysis; however,
as a specific illustration, we set $\delta_{V}$ = $-0.3$, $r_{V}$ = 308~Mpc, and $\Delta r$ = 18.46~Mpc.  This simplified representation is sufficient for our global analysis below. A more detailed $\chi^{2}$-fit analysis can be pursued later when more $K$-band data become available.  (However, we note that a more negative value
of $\delta_{V}$ than $-0.3$ would be disfavored by the SNe Ia data in the void region; see Figure~\ref{figDM}
below.)
Convergence to a $\Lambda$CDM model occurs in the 
$\Lambda$LTB inhomogeneous spacetime metric as $\delta(r)$ $\rightarrow$ 0. In our mass parameterization of the KBC LLV, the convergence to the universal mass density occurs at $\approx1.3\times$ the radius of the void.

We now introduce the LTB physical framework to determine dynamical observables within the void using a fully relativistic and nonlinear cosmological model.  The LTB inhomogeneous cosmological model has been extensively studied in the context of whether a GLV could explain the cosmic acceleration seen in SNe Ia (Garc\'{i}a-Bellido et al.\ 2008a, 2013b; Moss et al.\ 2011; Zibin \& Moss 2011; Bull et al.\ 2012; Wang et al.\ 2012; Iribarrem et al.\ 2014; Yan et al.\ 2015; Zhang et al.\ 2015; Chirinos Isidro et al.\ 2016); thus, we use the formalism developed in these studies. 

To start, we invoke the LTB inhomogeneous spacetime metric,
\begin{equation} 
ds^{2} = c^{2} dt^{2} - \frac{(R'(t,r))^{2}}{1 - k(r)} dr^{2} - R^{2}(t,r)d\Omega^{2} \,,
\label{eqLTB}
\end{equation}
where the prime denotes $\partial$/$\partial$r.  
We are modeling well into the era of Dark Energy domination; thus, we will describe our void dynamics both by the LTB void contribution and by pressure with a nonzero cosmological constant, otherwise known as the $\Lambda$LTB model.  We use the Friedmann Equation for the LTB metric,
\begin{equation} 
{1\over c^2}\dot{R}^{2} = -k(r) + \frac{2 M(r)}{R} + \frac{1}{3} \Lambda R^{2} \,,
\label{eqFE}
\end{equation}
to solve for $R(t, r)$.  The functions $M(r)$ and $k(r)$ are arbitrary functions of $r$, but physically, $M(r)$ can be interpreted as a generalized version of mass, and $k(r)$ as a generalized 
version of spatial curvature (Chirinos Isidro et al.\ 2016).  Within the KBC LLV, the scale factor, $R(t, r)$, is a function of both the radial coordinate and time.  External to the void, $R(t, r)$ in $\Lambda$LTB approaches the homogeneity limit of the $\Lambda$FRW spacetime metric with $R(t, r) = a(t)\times r$, where $a(t)$ is the usual scale factor.  

In order to describe dynamical observables within the LTB spacetime, we introduce, respectively, the line-of-sight Hubble and transverse parameters,
\begin{equation} 
H(t, r) = \frac{\dot{R}(t, r)}{R(t, r)}~~\text{and}~~\widetilde{H}(t, r) = \frac{\dot{R'}(t, r)}{R'(t, r)} \,,
\label{eqTrans}
\end{equation} 
where $H(t, r)$ corresponds to the observed Hubble parameter, and $\widetilde{H}(t, r)$ is relevant in the kSZ effect (Enqvist \& Mattsson 2007; Blomqvist \& M{\"o}rtsell 2010; see Section~\ref{seckSZ}).  Equation~\ref{eqFE} can now be rewritten in terms of the conventional fractional energy densities (see, e.g., Sundell et al.\ 2015),
\begin{equation} 
\Omega_{M}(r) = \frac{2 M(r) c^2}{H^{2}_{0}(r) R_{0}^{3}(r)},~~\Omega_{k}(r) = \frac{-k(r)c^2}{H^{2}_{0}(r) R_{0}^{2}(r)},~~\text{and}~~\Omega_{\Lambda} = \frac{\Lambda c^{4}}{3 H^{2}_{0}(r)} \,,
\label{eqOMr}
\end{equation} 
where a subscript 0 denotes the value of the observable evaluated at the present time, $t_{0}$. Thus, we obtain
\begin{subequations} 
\begin{align}
\biggr(\frac{\dot{R}}{R} \biggl)^{2} &= H^{2}_{0}(r) E^{2}(R, r) \,, \label{eqe2a} \\
E^{2}(R, r) &\equiv \biggr[\Omega_{M}(r)~\biggr(\frac{R_{0}(r)}{R(t, r)}\biggl)^{3} +~\Omega_{k}(r)~\biggr(\frac{R_{0}(r)}{R(t, r)}\biggl)^{2} +~\Omega_{\Lambda}(r) \biggl]\,.
\label{eqe2b}
\end{align}
\end{subequations}

We again adopt the labels ``in" and ``out", respectively, for the cosmological parameters internal and external to the LLV.
We now define $\delta(r)$ through the equation
\begin{equation}
\Omega_M(r) = \Omega_{M,out} [1 + \delta(r)] \,.
\end{equation}
Assuming $\Omega_\Lambda$ is universal, we get
\begin{equation}
\Omega_\Lambda =1 - \Omega_{M,out} \,.
\end{equation}
Using the above two equations together with 
\begin{equation}
\Omega_{M,in} + \Omega_{k,in} + \Omega_{\Lambda} = 1 \,,
\end{equation}
we can now express the first two fractional energy densities in Equation~\ref{eqOMr} 
in terms of the mass density profile, $\delta(r)$, 
which goes to zero at large $r$ (where $\Omega_M(r)=\Omega_{M,out}$), as
\begin{subequations} 
\begin{align}
2M(r) c^2 &= H^{2}_{0}(r) R^{3}_{0}(r)~\Omega_{M,out}~[1 + \delta(r)] \,, \label{eqMr} \\
k(r) c^2 &= H^{2}_{0}(r)R^{2}_{0}(r)~\Omega_{M,out}~\delta(r) \label{eqkr} \,.
\end{align}
\end{subequations} 
It is relevant to note that the curvature, $k(r)$, is fundamentally related to the mass density profile, $\delta(r)$.  Since $\delta(r) <0$, then from Equation~\ref{eqkr}, $k(r) < 0$. The compensation of a positive curvature energy density (which corresponds to a negative $k(r)$) for a matter underdensity is a {\em natural\/} consequence of the $\Lambda$LTB model. 

The radial null geodesics ($ds^2=0$) of the LTB inhomogeneous spacetime are given by
\begin{equation} 
\frac{dt}{dr} = -\frac{R^\prime(t,r)}{c\sqrt{1 - k(r)}} \,.
\label{eqdtdr}
\end{equation}
Also, the relation of the radial variable, $r$, to the redshift, $z$, is given by
Alnes \& Amarzguioui (2006) and Enqvist \& Mattsson (2007) as
\begin{equation} 
\frac{1}{1 + z}\frac{dz}{dr} = \frac{\dot{R}^\prime(t,r)}{c\sqrt{1 - k(r)}} \,.
\label{eqdzdr}
\end{equation}
We examine numerical solutions to Equation~\ref{eqFE} in order to solve for $R(t,r)$.  
Substituting Equation~\ref{eqe2a} into Equation~\ref{eqFE}, we separate 
variables and integrate to obtain
\begin{equation} 
c(t - t_{B}(r)) = \frac{1}{H_{0} (r)} \int_{0}^{R} \frac{d \tilde{R}}{\tilde{R}~E(\tilde{R}, r)} \,,
\label{eqc}
\end{equation}
where $t_B$ denotes the Big Bang time.
We can now numerically solve Equation~\ref{eqc} for $R(t,r)$.  
We work in the usual gauge $R_{0}(r) = r$ (see Enqvist 2008; Wang \& Zhang 2012; Zhang et al.\ 2015)
and use our parameterization of $\delta(r)$ in Equations~\ref{eqMr}, \ref{eqkr}, \ref{eqOMr},
and \ref{eqe2b}.
At the present time, $t=t_{0}$, we obtain a numerical solution for $t_{B}(r) = r$ from Equation~\ref{eqc}.  
We then substitute this $t_B(r)$ into Equation~\ref{eqc} at an arbitrary time $t$. Finally, we
solve Equation~\ref{eqc} by numerical integration and subsequent inversion to get $R(t,r)$.
The formula for the luminosity distance within the LTB spacetime is then
\begin{equation} 
D_{L, \Lambda LTB} (z)  = (1+z)^{2} R(t(z), r(z)) \,.
\label{eqLTBdL}
\end{equation} 

\begin{figure*}
\hskip -0.15cm
\centerline{\includegraphics[width=8.4cm,angle=0]{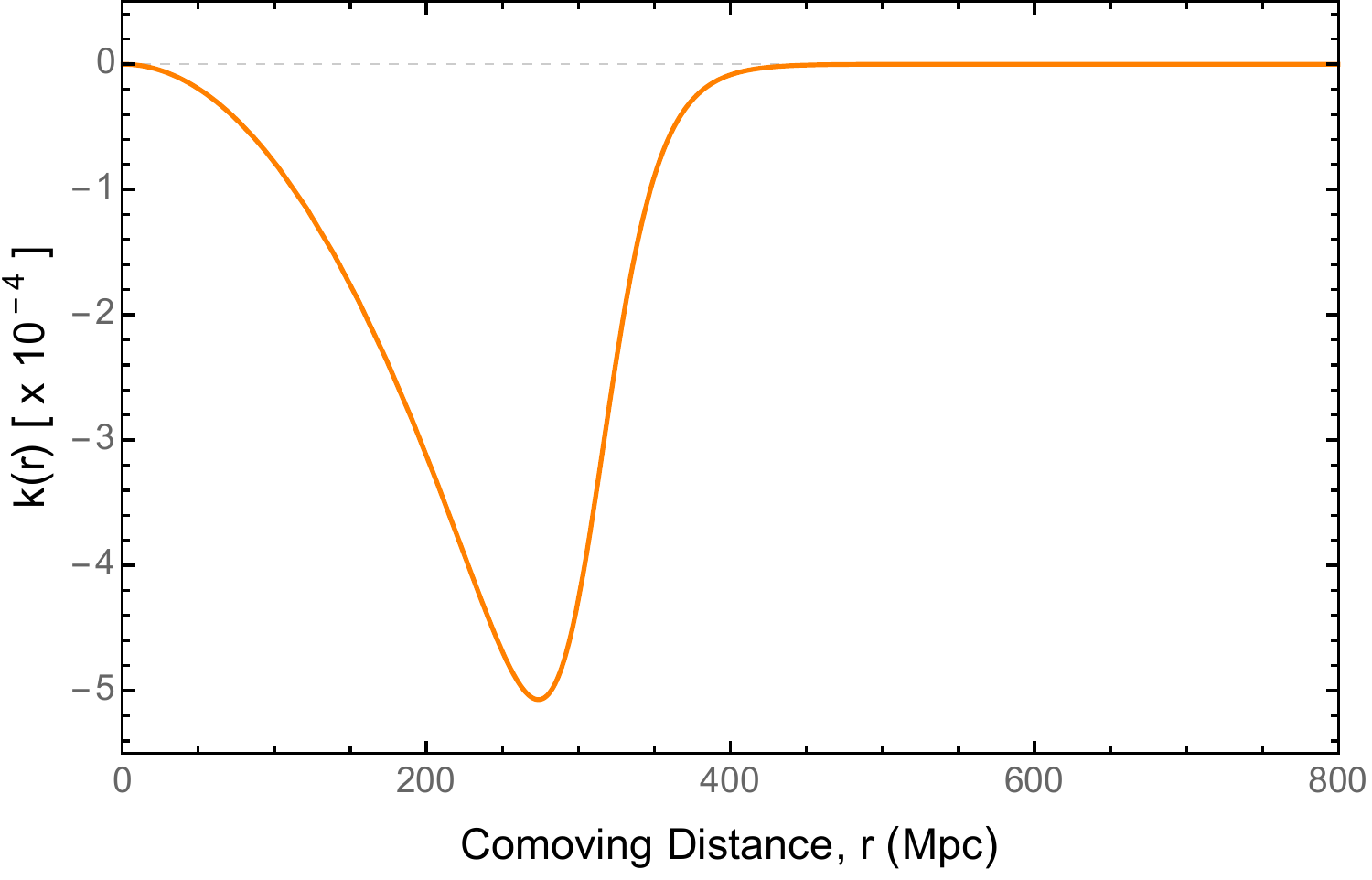}
\includegraphics[width=8.6cm,angle=0]{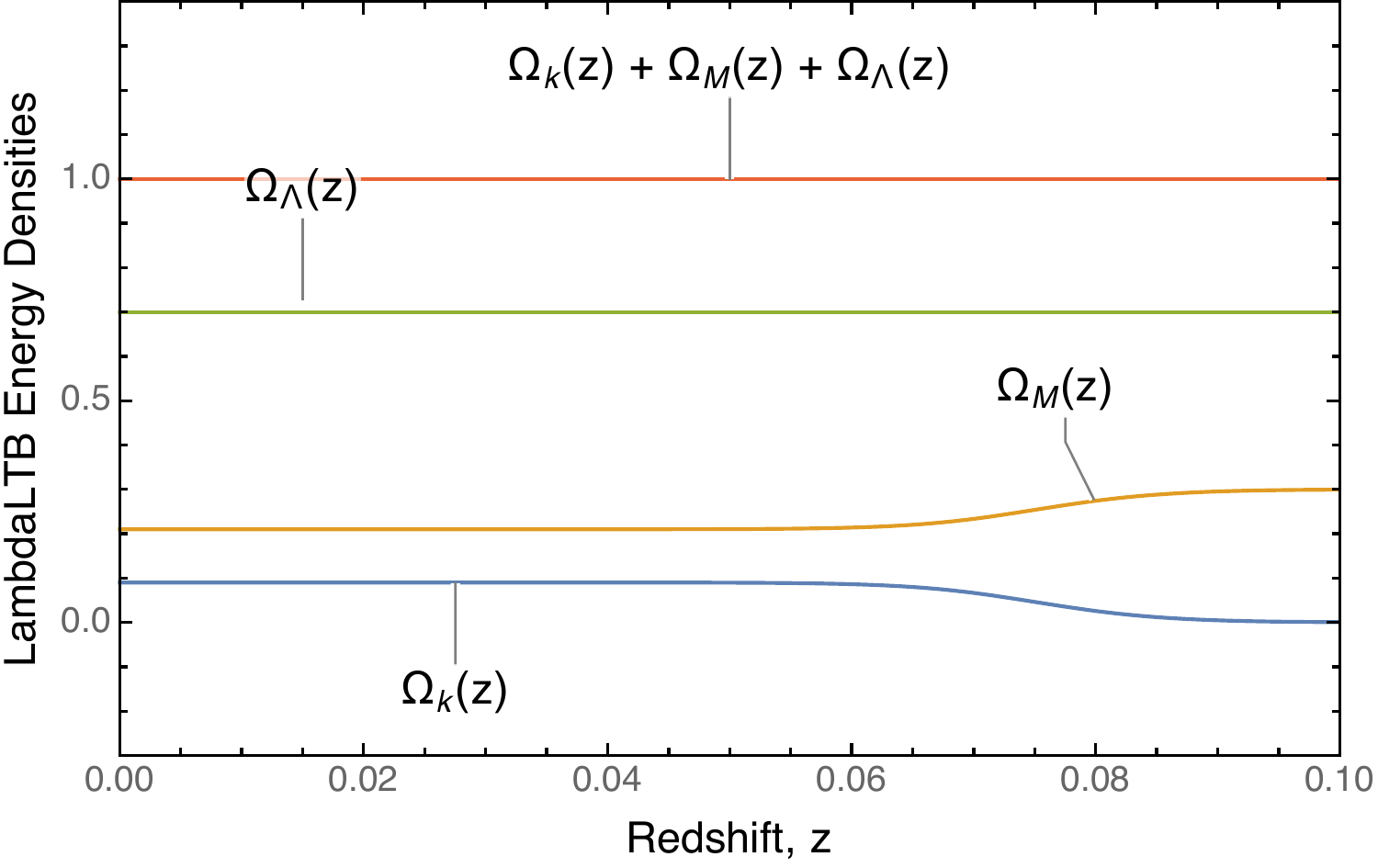}}
\caption{(a)  
$k(r)$ vs. comoving distance, $r$, for the mass density parameterization of Figure~\ref{figData}.
$k(r)$ can be interpreted as a generalized version of spatial curvature. Its functional deviation from zero 
directly translates into perturbed dynamical cosmological observables in the $\Lambda$LTB spacetime.  
(b) 
Energy densities for the $\Lambda$LTB parameters in units of the $\Lambda$LTB critical density vs. redshift.
On redshift scales $0 < z \lesssim 0.06$, $\Omega_{M}(r) = 0.21$ and $\Omega_{k}(r) = 0.09$, while on redshift scales $z \gtrsim 0.09$,  $\Omega_{M}(r) = 0.3$ and $\Omega_{k}(r) = 0$.  $\Omega_{\Lambda}(r)$ is assumed to be constant, $\Omega_{\Lambda}(r)=0.7$.
}
\label{figFRW}
\vskip 0.5cm
\end{figure*}

In Figure~\ref{figFRW}(a), we plot $k(r)$ versus $r$ for the mass density parameterization of Figure~\ref{figData}. 
In Figure~\ref{figFRW}(b), we show the energy densities for the $\Lambda$LTB parameters
in units of the $\Lambda$LTB critical density versus redshift.  The curvature density (blue curve) goes to zero outside the void as the mass density (orange curve) increases.  Both the $\Lambda$ density (0.7) and the sum of the three energy densities are taken to be constant at all redshifts.

Within this theoretical framework, we can now describe the redshift dependence of any dynamical cosmological observable within the KBC LLV.  Cosmological observables in an inhomogeneous cosmological model are different than in a homogeneous cosmological model, and this is a direct result of the underlying KBC LLV mass density distribution.  These dynamical cosmological observables are necessary for a description of the luminosity distance-redshift relation and the linear kSZ effect under the radial inhomogeneity of the KBC LLV.

\section{Measurements of $H_{0}$}
\label{secSNe}

In this section, we discuss the tension in measurements of the Hubble constant, $H_{0}$, in light of our results.  
We refer to the different redshift ranges used for the measurement of $H_{0}$ as ``local", ``intermediate", 
and ``cosmic".  For the latest ``local" measurements of the Hubble constant, R16 used distance-ladder-based  
techniques: (i) three absolute distance anchors, including the megamaser host NGC 4258, 
(ii) the Leavitt period$-$luminosity relation for Cepheid variable stars, and (iii) the empirical relationship between 
the observed light curve and the intrinsic luminosity of SNe Ia.  
They found a value of $H_{0} = 73.24 \pm 1.74~\rm{km~s^{-1}~Mpc^{-1}}$.  
For the latest ``Intermediate" measurements, Addison et al.\ (2017) reported a value of 
$H_{0} = 66.98 \pm 1.18~\rm{km~s^{-1}~Mpc^{-1}}$ from combining galaxy and 
Lyman $\alpha$ (Ly$\alpha$) baryon acoustic oscillations (BAO) with an estimate of the primordial deuterium abundance.  
In addition, Bonvin et al.\ (2017) reported a value of $H_{0} = 71.9^{+2.4}_{-3.0}~\rm{km~s^{-1}~Mpc^{-1}}$ 
from their analysis of three imaged quasar systems using gravitational time delays.
For the latest ``cosmic" measurements, Ade et al.\ (2014) reported a value of $H_{0} = 66.93 \pm 0.62~\rm{km~s^{-1}~Mpc^{-1}}$
from the \textit{Planck} CMB measurements and a standard $\Lambda$CDM cosmology.
This is similar to the value of $H_{0} = 69.3 \pm 0.7~\rm{km~s^{-1}~Mpc^{-1}}$ found by Bennett et al.\ (2014) by
combining data from WMAP9, the SPT, and the ACT
with BAO measurements from the Baryon Oscillation Spectroscopic Survey (BOSS).  
Overall, in $\Lambda$CDM, higher $H_{0}$ is preferred by the distance ladder, and lower $H_{0}$ is preferred by the CMB data.
BAO data are in accord with the $H_{0}$ from the CMB.

As alluded to in R16, the tension in $H_{0}$ could be due to systematic errors at high redshifts.  
Alternatively, a new radiation degree of freedom ($\Delta$$N_{\text{eff}}< 0.5$) or decaying dark matter are 
possible ways that new physics can modify high-redshift observations without any direct bearing on our analysis.  Various groups have also recently reanalyzed the relevant systematics associated with distance ladder based measurement data (Freedman 2017).  Motivated by the fact that the calibration of SNe~Ia has the possibility of being biased by human analysis, Zhang et al.\ (2017) performed a blinded determination of $H_{0}$ from the Riess et al.\ (2011) and Efstathiou (2014) SNe Ia dataset. They reported a higher total relative uncertainty of 4.4\% as compared to the total relative uncertainty of 3.3\% determined in 
Riess et al.\ (2011).  In addition, Feeney, Mortlock, \& Dalmasso (2017) sought to more accurately describe the tension in $H_{0}$ by including the tails of the likelihoods (modeling the outliers and eliminating arbitrary cuts in the data analysis) in the two separate parameter estimations through a Bayesian hierarchical model (BHM).  After performing their statistical reanalysis of the R16 SNe Ia data, they found a value $H_{0} = 73.15 \pm 1.78~\rm{km~s^{-1}~Mpc^{-1}}$ when SNe Ia outliers are reintroduced into the R16 dataset. They concluded that the tension in $H_{0}$ does indeed persist.  Furthermore, Follin \& Knox (2017) investigated the possibility that the tension in $H_{0}$ could be due to systematic bias or uncertainty in the Cepheid calibration step of the distance ladder measurement by R16; they concluded it could not.  A similar conclusion was reached in Dhawan et al. (2018) with respect to the possibility of SNe Ia systematics arising from variations between optical and NIR wavelengths, such as dust extinction.  Most recently, the local Dark Energy Survey Year 1 (DES Y1) Results (Abbott et al.\ 2017) suggest that the discrepancy in measurements of the Hubble constant are not as big as previously claimed.  Their combined analysis of the DES Y1, Joint Lightcurve Analysis (JLA), BAO, and \textit{Planck} datasets led to a value of $H_{0} = 68.2 \pm 0.6~\rm{km~s^{-1}~Mpc^{-1}}$.

\subsection{Low-redshift Supernovae Type Ia}

SNe Ia can be calibrated through their light curves to be standard candles and thus are used to determine 
cosmological parameters.  In a $\Lambda$FRW cosmology, the luminosity distance, $D_{L}(z)$, is given by
\begin{equation} 
D_{L}(z) = \frac{c}{H_{0} (1 + z)} \int_{0}^{z} \frac{dz'}{E(z')}  \,,
\end{equation}
where $E(z) = \sqrt{\Omega_{M} (1 + z)^{3} + \Omega_{k} (1 + z)^{2} + \Omega_{\Lambda}}$.  
At low redshifts, a Taylor expansion gives
\begin{equation} 
D_{L}(z) = \frac{c z}{H_{0}} \biggr\{1 + \frac{1}{2} (1 - q_{0})z - \frac{1}{6} (1 - q_{0} - 3 q_{0}^2 + j_{0}) z^{2} + O(z^{3}) \biggl\}   \,,
\end{equation}
where $q_{0}$ and $j_{0}$ are the deceleration and jerk parameters, respectively.  
Using measurements of 740 SNe Ia at z $\lesssim 1$ (Betoule et al.\ 2014), we determine 
values of $q_{0} = -0.55$ and $j_{0} = 1$.
The $q_{0}$ and $j_{0}$ uncertainties only effect the Hubble constant determination at the $0.1\%$ level.  
We note that R16 introduces a model dependence into their analysis by assuming a constant redshift 
dependence in $q_{0}$ and $j_{0}$.  Strictly speaking, to do the analysis properly in the presence of the KBC LLV, 
we need to take into account the variation of $q_{0}$ and $j_{0}$ with the local density inhomogeneities.  
For the KBC LLV parameterization of Figure~\ref{figData} with $\delta_{V}$ = $-0.3$, we find $q_{0} = - 0.536$ 
and $j_{0} = 0.835$ within the void; this change in $q_{0}$ and $j_{0}$ is at or below the $0.1 \%$ level in $H_{0}$.

The distance modulus,
\begin{equation} 
\mu(z) = m - M \,,
\end{equation}
is related to the luminosity distance by
\begin{equation} 
\mu (z)  = 5 \log_{10} \biggr(\frac{D_{L} (z)}{{\rm Mpc}} \biggl)  +  25 \,.
\end{equation}
Here, $m$ is the apparent magnitude and $M$ is the absolute magnitude in a given color band. 
Since the $B$ band is where the most recent measurements of $H_0$ have been made, we use the subscript $B$ hereafter.
Selective absorption is corrected for through the use of colors and a reddening law, after which the
magnitudes are usually denoted by $m_B^{0}$ and $M_B^{0}$.  At low redshifts, the SNe Ia distance modulus can be written
\begin{equation} 
m_{B}^{0} - M_{B}^{0} = 5~\text{log}~ \frac{c z}{H_{0}} \biggr\{1 + \frac{1}{2} (1 - q_{0}) z - \frac{1}{6} (1 - q_{0} - 3 q_{0}^2 + j_{0}) z^{2} \biggl\}+25 \,.
\label{eqdistmod}
\end{equation}
This equation may be re-expressed as
\begin{equation} 
\text{log}~c z \biggr\{1 + \frac{1}{2} (1 - q_{0}) z - \frac{1}{6} (1 - q_{0} - 3 q_{0}^2 + j_{0}) z^{2} \biggl\} - 0.2 m_{B}^{0} = \text{log}~H_{0} - 0.2 M_{B}^{0}-5 \,.
\end{equation}
It is then convenient to define a quantity (R16)
\begin{equation} 
a_{B} = \text{log}~cz \biggr\{1 + \frac{1}{2} (1 - q_{0}) z - \frac{1}{6} (1 - q_{0} - 3 q_{0}^2 + j_{0}) z^{2} \biggl\} - 0.2 m_{B}^{0} \,,
\label{aB}
\end{equation}
which is solely determined by the measured redshifts and apparent magnitudes of the SNe Ia.  

In the following analysis, we use the R16 dataset, which was provided to us by A. Riess (private communication).
It includes 217 SNe Ia in the redshift range $0.0233<z<0.15$, which is the primary redshift range used 
by R16 for determining the most precise local measurement of $H_{0}$.  
The values given to us were $z$, $m_{B}^{0}$, and total errors in $m_{B}^{0}$.  
This SNe Ia dataset is the result of a series of 
``quality cuts" applied to the Supercal SNe Ia dataset of Scolnic et al.\ (2015).  
In particular, SNe Ia outliers ($>3\sigma$) for the Hubble diagram intercept, 
$a_{B}$, have been excluded (cutting out 3\% of the sample), as have
SNe Ia data $z<0.0233$, in order to avoid the possible influence of coherent bulk flows.

\begin{figure*}
\hskip -0.15cm
\centering
\includegraphics[width=13cm,angle=0]{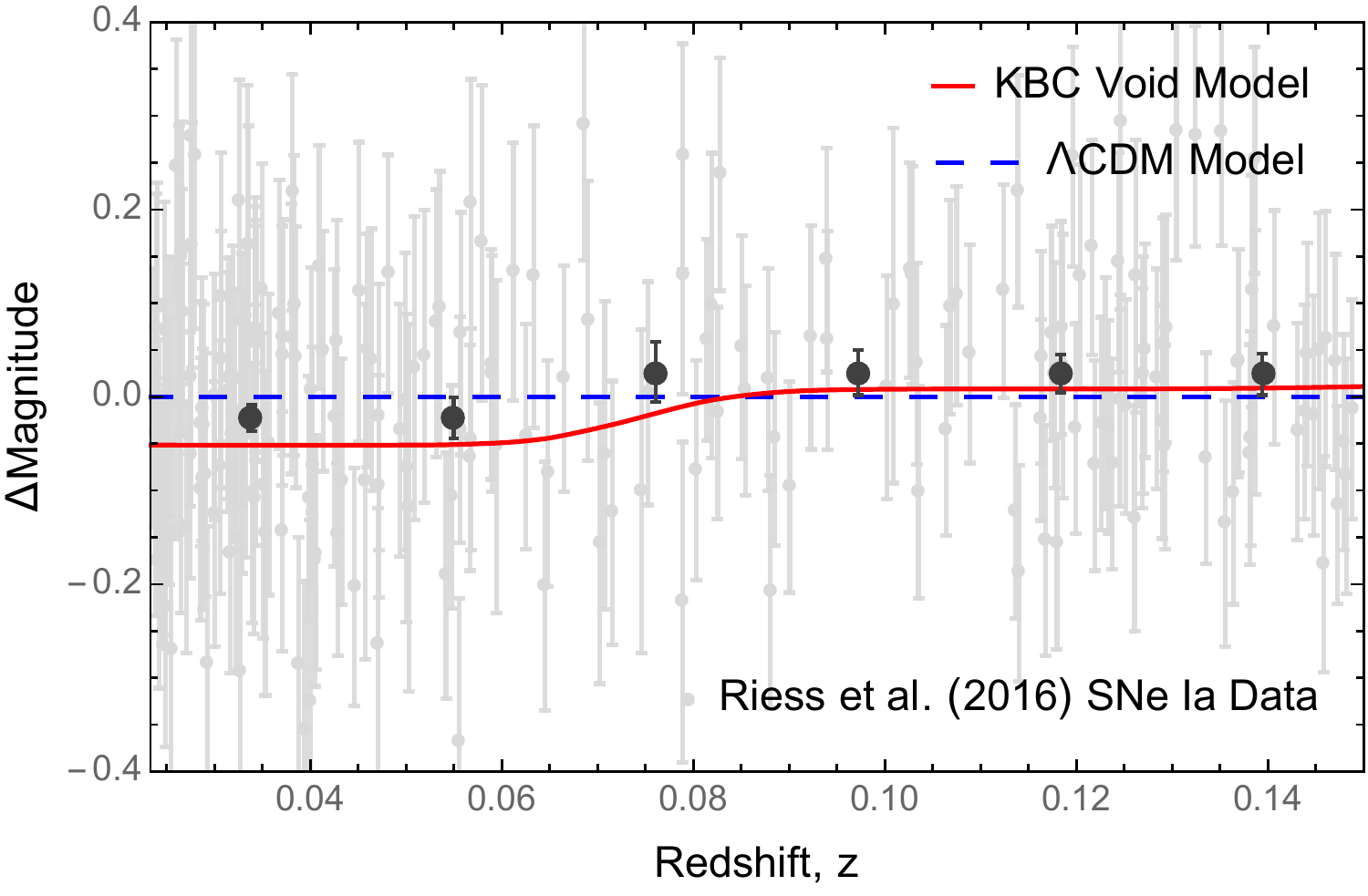}
\caption{Distance modulus differences vs. redshift between the standard $\Lambda$CDM cosmological model (blue dashed) 
and the $\Lambda$LTB model for the KBC LLV (red solid) over the redshift range $0.0233<z<0.15$ 
(assuming $H_{0}$ = 73.24 $\rm{km~s^{-1}~Mpc^{-1}}$). 
We also show the distance modulus differences between the predictions of $\Lambda$CDM and the 
measured values of individual SNe Ia (A. Riess, private communication) (gray small points) vs. redshift. 
We show the weighted means (black large circles) for the SNe Ia data in redshift bins of 0.021.  Both models appear as equally good fits to the SNe~Ia data.  
\label{figDM}
}
\vskip 0.5cm
\end{figure*}

In Figure~\ref{figDM}, we plot the distance modulus differences, $\Delta$Magnitude, between our KBC LLV 
parameterization (red curve) and the standard $\Lambda$CDM cosmological model 
(assuming $H_{0}$ = 73.24~$\rm{km~s^{-1}~Mpc^{-1}}$) (blue dashed line) vs. redshift.
R16 found $M_{B}^{0} = -19.25$ with an uncertainty of less than $0.1$ magnitude for their full
$0.0233<z<0.15$ SNe~Ia sample, but here we use the value that we determined after considering only 
the SNe~Ia data at $z > 0.07$, $M_{B}^{0} = -19.24$.
We also show the distance modulus differences between the 
measured values of individual SNe~Ia (gray small circles) and the predictions of $\Lambda$CDM.
The weighted means (black large circles) visually suggest that an 
upward shift occurs at the edge of the void, at $z = 0.07$.  We note that for both models, 
$\chi^{2}/N_{data}\sim$0.9. There is clearly too much scatter in the individual SNe~Ia measurements to 
rule out the KBC LLV.
We conclude that the SNe~Ia data neither reject nor sensitively probe the existence of the KBC LLV.

A linear fit of the weighted SNe Ia apparent magnitudes determines $a_{B}$ (R16), after which
$H_{0}$ can be determined from
\begin{equation} 
\text{log} \; H_{0} = \frac{M_{B}^{0} + 5 a_{B} + 25}{5}  \,.
\label{eqsolveHo}
\end{equation}
In order to investigate the effects of the KBC LLV underdensity, we take $z = 0.07$ to define the edge 
of the void.  We then calculate the $a_{B}$ values for two redshift 
intervals: inside ($0.0233 < z < 0.07$) and outside ($0.07 < z < 0.15$) the void, finding 
\begin{equation}
a_{B, in} = 0.716475 \pm 0.00248~~\text{and}~~a_{B,out} = 0.708909 \pm 0.00251.  
\end{equation}
This corresponds to a $2.14 \sigma$ downward shift in $a_{B}$ from inside to outside the void.

\subsection{Effects of Local Density Contrast and Curvature}

We now describe our modeling of the effects of local density contrast and curvature.  
In $\Lambda$CDM, $H^{\Lambda CDM}(z)$ is the background Hubble parameter, given by
\begin{equation} 
H^{\Lambda CDM}(z) = H_{0}^{out} \sqrt{\Omega_{M} (1 + z)^{3} + \Omega_{\Lambda}} \,.
\end{equation}
Using the R16 SNe~Ia dataset over the redshift range $0.07<z<0.15$  and Equation~\ref{eqsolveHo},
we obtain $H_{0}^{out} = 72.26 \pm 0.42~\rm{km~s^{-1}~Mpc^{-1}}$.
Within the void, we use the $\Lambda$LTB prediction of $H(t(z), r(z))$ from the solution to 
Equation~\ref{eqFE} (Enqvist 2008), which approaches the $\Lambda$CDM result at high redshift.
The quantity of interest is the Hubble parameter difference
\begin{equation} 
\Delta H(z) = H^{\Lambda LTB}(z) - H^{\Lambda CDM}(z) \,.
\label{DHz}
\end{equation}

\begin{figure*}
\centering
\includegraphics[width=13cm,angle=0]{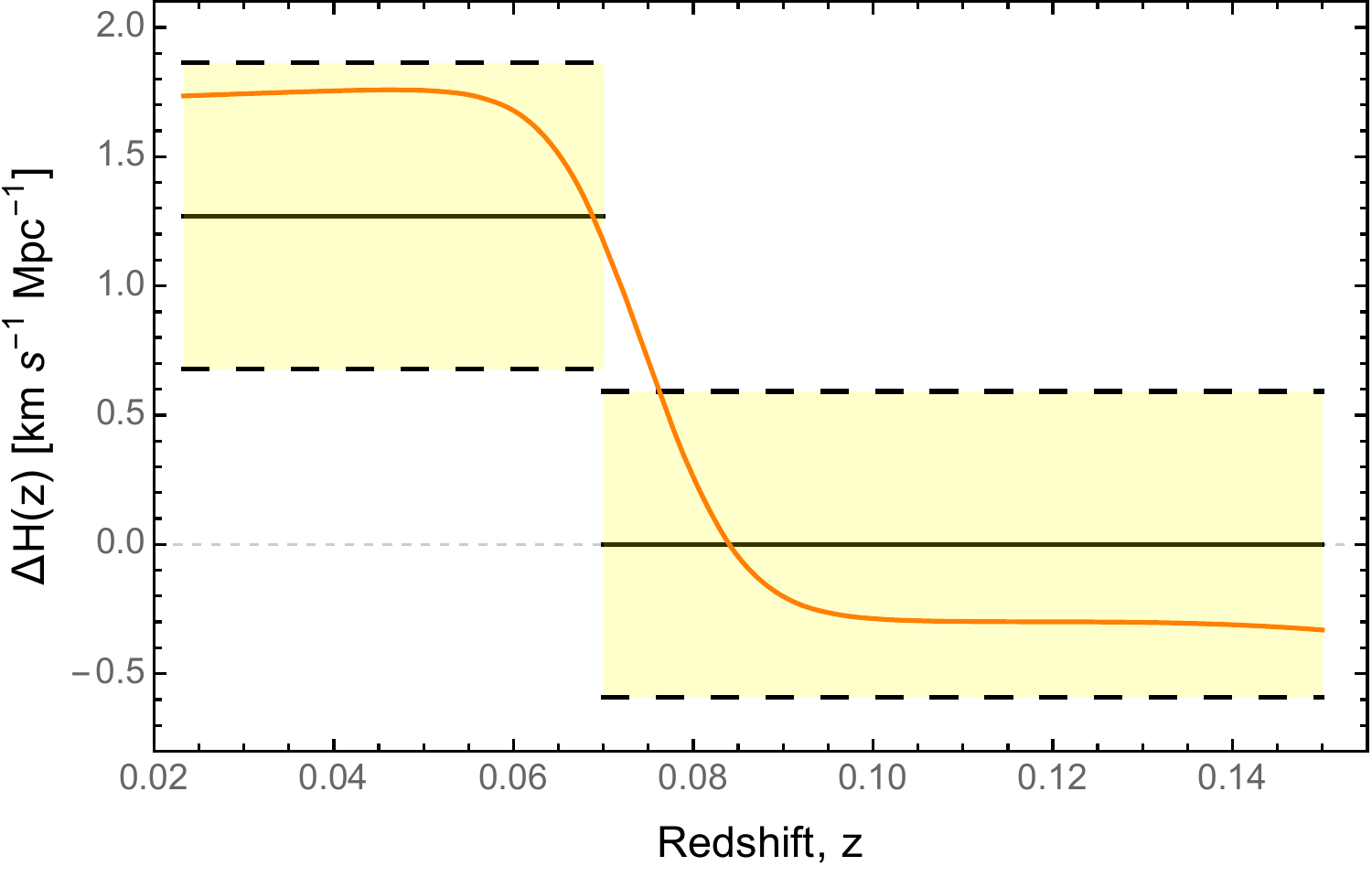}
\caption{
Dependence of $\Delta H(z)$, as defined by Equation~\ref{DHz}, on redshift. 
The orange curve represents the $\Lambda$LTB
model prediction with the KBC LLV density parameterization of Figure~\ref{figData}.
The yellow bands are observational results from the R16 SNe~Ia data 
with $1\sigma$ uncertainties for the two redshift ranges.
\label{H0out}
}
\end{figure*}

The smooth orange curve in Figure~\ref{H0out} shows the model prediction. The dependence of $\Delta H(z)$
on $z$ is relatively flat both within and outside the void, with a sharp decline at the void interface. The 
yellow bands represent observational values from the R16 SNe Ia dataset when separated into the two
redshift ranges $z=0.02-0.07$ and $z=0.07-0.15$. The bands show $1\sigma$ statistical errors on
$\Delta H$. The mean $\Delta H$ values are obtained from the weighted means of $a_B$ over the
respective redshift ranges with $M_B^0=-19.24$; the obtained values of $\Delta H$ are insensitive to
changes in $M_{B}^{0}$ of $\pm 0.01$.

From the above, we conclude that the R16 SNe~Ia data do not exclude the KBC LLV. Indeed, the R16
data show evidence for a sharp decline in the Hubble constant at $z=0.07$ that corresponds to the void 
boundary. The drop in $H_0$ at the edge of the void then reduces the discrepancy with the \textit{Planck} $H_0$
from $3.4\sigma$ to $2.75\sigma$.  This amount of relaxation of the Hubble constant tension is in good agreement with the numerical result obtained by Wu \& Huterer (2017) for a local underdensity of $\delta = -0.3$.  They derive their numerical predictions from the formalism in Marra et al.\ (2013), where a $\Lambda$LTB model was used to compute a nonlinear correction to the impact of local density contrast on local deviations of the Hubble constant. The conclusion of their numerical study is that an underdensity sufficient to relax fully
the $H_0$ tension is extremely unlikely in $\Lambda$CDM.

\section{Linear kSZ Effect}
\label{seckSZ}
We now discuss the linear kSZ effect produced by the radial inhomogeneity introduced by the KBC LLV.  
We draw on the pioneering theoretical studies of Garc\'{i}a-Bellido et al.\ (2008a), Moss et al.\ (2011),
Zhang et al.\ (2011, 2015), Zibin \& Moss (2011), and Bull et al.\ (2012).
We start by noting that the void-induced linear kSZ power in direction $\hat{n}$ is given by 
\begin{equation} 
\Delta T_{kSZ} (\hat{n}) = T_{CMB} \int_{0}^{z_{e}} \delta_{e}(\hat{n},z) \frac{\overrightarrow{V_{H}}(\hat{n},z) \cdot \hat{n}}{c}  d\tau_{e} \,.
\label{linearkSZ}
\end{equation}
Here $T_{CMB}$ = 2.73~K is the average temperature of the primary CMB anisotropies,
$\delta_{e}$ is the density contrast of electrons, and $\tau_e$ is the optical depth along the line of sight,.
As noted by Zhang et al.\ (2015), if one integrates far enough away from the relevant void model edge, 
then the integration result does not change; we choose $z_e=100$ for the upper limit of the integration.
Equation~\ref{linearkSZ} is referred to as the \textit{linear} kSZ effect, because it is dependent upon the first-order void 
induced ``bulk flow'' term. It does not account for nonlinear perturbation terms, such as the standard dipole 
peculiar-velocity perturbations in $\Lambda$FRW spacetime metric models (Zibin \& Moss 2011). 

Generally, $V_H$ is due to both Doppler and Sachs-Wolfe anisotropies generated by the void. However,
the Sachs-Wolfe component of $V_H$ is negligible in comparison to the Doppler component for $z_{edge} < 1$
(Figure~2 of Caldwell et al.\ 2008).  Since the scale of the KBC LLV is $z_{edge} < 0.1$, and since
the void is not extremely deep in the matter density parameterization at $z = 0$ ($| {V_H} |\ll c$),
we may approximate $V_H$ as the non-relativistic CMB Doppler anisotropy.
Thus, following Zhang et al.\ (2015), we make the assumption that 
\begin{equation} 
V_{H} \simeq [\widetilde{H}(t(z), r(z)) - \widetilde{H}(t(z), r(z_{e}))] R(t(z), r(z)) \,.
\label{eqVH}
\end{equation}

In order to integrate over redshift in Equation~\ref{linearkSZ}, we follow Zibin \& Moss (2011) and Moss et al.\ (2011) 
and parameterize the optical depth along the line of sight, $\tau_{e}(\hat{n})$, as 
\begin{equation} 
\frac{d\tau_{e}}{dz} = \sigma_{T} n_{e}(z) c \frac{dt}{dz} = \frac{\sigma_{T} \theta^{2} f_{b} (1 - Y_{He}/2) \Omega_{M} (1 + \delta(r(z)))}{24 \pi G m_{p}} c \frac{dt}{dz} \,,
\end{equation}
where $\sigma_{T}$ is the Thomson cross section, $f_{b}$ = 0.168 is the baryon fraction as implied by CMB observations 
(Larson et al.\ 2011), $Y_{He}$ = 0.24 is the helium mass fraction, $m_{p}$ is the mass of the proton, and $\theta$ is given by 
\begin{equation} 
\theta = (\widetilde{H} + 2 H) \,.
\end{equation}
Again following Zibin \& Moss 2011, we invoke the Limber (1953) approximation to simplify the $\delta_{e}$ term,
\begin{equation} 
C_{\ell} \simeq \frac{16 \pi^{2}}{(2 \ell + 1)^{3}} \int_{0}^{z_{e}} dz \frac{dr}{dz} r(z) F^{2}(r(z)) P_{\delta}\biggl(\frac{2 \ell + 1}{2 r(z)},z \biggr)  \\,.
\label{power}
\end{equation}
In this case, $C_{\ell}$ is the linear kSZ power at multipole $\ell$, and
\begin{equation} 
F(r) \equiv \frac{V_{H}(r)}{c} \frac{d\tau_{e}}{dz} \frac{dz}{dr} \,.
\end{equation}
Similarly, we describe the $\Lambda$FRW matter-power spectrum component of Equation~\ref{power}, $P_{\delta}(k, z)$, 
using the publicly available and widely-accepted CAMB code, where we have made use of the Halofit feature to physically 
realize the kSZ power introduced on relevant nonlinear scales (Lewis et al.\ 2000).  
In addition, we define the conventionally scaled quantity 
\begin{equation} 
D_{\ell} \equiv \frac{\ell (\ell + 1) C_{\ell} }{2 \pi} \,.
\end{equation}
In Section~\ref{secConstraints}, we will make use of the quantity $T^{2}_{CMB}D_{3000}$ when considering the linear kSZ power introduced by the KBC LLV.

\section{CMB Constraints on the KBC LLV}
\label{secConstraints}
\begin{figure*}
\hskip 2.5cm
\hskip -2.6cm\includegraphics[width=16cm,angle=0]{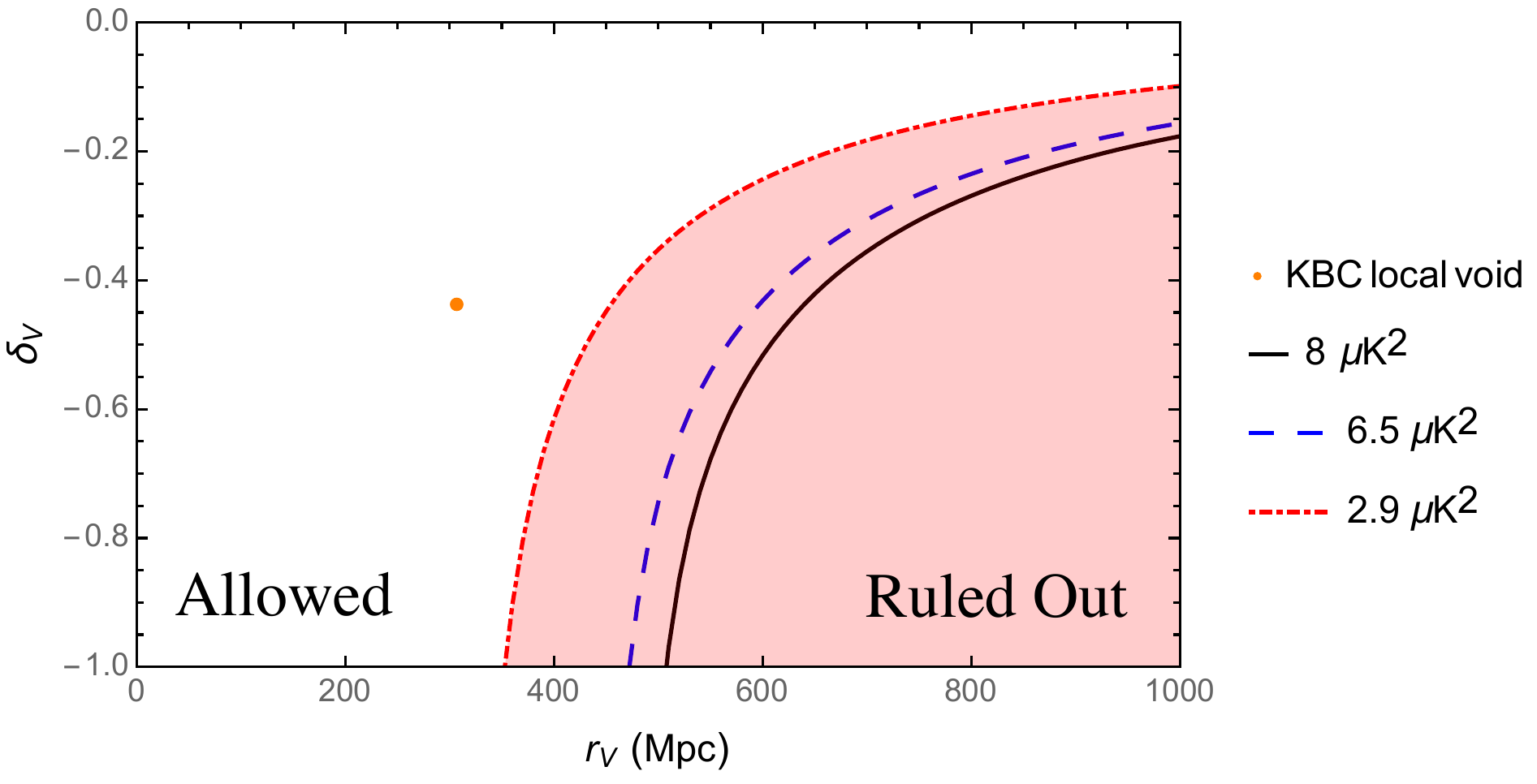}
\caption{Contours in CGBH void model parameter space that correspond to constant values 
of $T^{2}_{CMB}D_{3000}$ in $\mu K^{2}$; we have set the values to be equal to the constraints 
given in Section~\ref{secConstraints}.  
The point corresponding to ($r_{V}$, $\delta_{V}$) = (308, $-0.3$) physically represents our mass density parameterization 
of the KBC LLV.  Note that this point lies in the region of parameter space that is fully consistent with the linear kSZ constraints.  
We use $H_{0}$ = 73.2 km {$\rm s^{-1}~\rm Mpc^{-1}$}.}
\label{kSZCons}
\vskip 0.5cm
\end{figure*}

We now compare the kSZ constraints from the SPT and the ACT with predictions from the $\Lambda$LTB model for our parameterization of the KBC LLV.
The 2008 ACT and 2008 SPT constraints correspond to a 95\% confidence upper limit on the CMB secondary anisotropy 
power; they are $< 8~\mu K^{2}$ (Das et al.\ 2011) and $<6.5~\mu K^{2}$ (Shirokoff et al.\ 2011), respectively.  
The 2014 and 2015 SPT constraints correspond to a 98.1\% confidence upper limit; they are $<2.9\pm 1.6~\mu {\rm K}^{2}$ 
(Crawford et al.\ 2014) and 
$<2.9 \pm 1.3~\mu {\rm K}^{2}$ (George et al.\ 2015), respectively.  
All constraints are measured at $\ell = 3000$ and at a frequency of $\sim150$~GHz.
In Figure~\ref{kSZCons}, we plot contours in CGBH void model parameter space that correspond to constant values of 
$T^{2}_{CMB}D_{3000}$ in $\mu K^{2}$; we have set the values to be equal to the above constraints. 
The point in the parameter space corresponding to 
($r_{V}$, $\delta_{V}$) = (308, $-0.3$) physically represents our parameterization of the KBC LLV (Figure~\ref{figData}).  
Previous studies (Zhang et al.\ 2015; Zibin \& Moss 2011; Moss et al.\ 2011; Ade et al.\ 2014) used these constraints to 
rule out a class of GLV models that fit the SNe~Ia data (and hence could have explained cosmic acceleration). 
The improved SPT kSZ constraints are more stringent than those used in these previous studies, but they are still 
fully compatible with the existence of the KBC LLV.

Although Zibin \& Moss (2011) found that the linear kSZ power for void models is insensitive to the value of $H_{0}$,
there are several other assumptions made in the modeling of the KBC LLV and associated dynamics that
affect Figure~\ref{kSZCons}.  First, there is uncertainty in the $K$-band luminosity density measurements shown in Figure~\ref{figData}.  This allows for a marginal amount of freedom when parameterizing the
KBC LLV.  For future work, it would be interesting to analyze how mass density parameterizations other than the CGBH model, 
such as the Gaussian spatial curvature function used in Romano et al.\ (2015) and Szapudi et al.\ (2014), would influence the 
stringency of the kSZ constraints on the KBC LLV.  Second, Figure~\ref{figData} is sensitive to the modeling of the void dynamics 
(e.g., see the Zibin \& Moss 2011 comparison of the ``Hubble Bubble'' FRW perturbation void model in Zhang et al.\ 2011 with 
the LTB void model in Zibin \& Moss 2011).  
Finally, Figure~\ref{figData} is dependent on how one chooses to model the linear kSZ first-order void induced ``bulk flow'' 
monopole term model, 
$V_{H}(r)$.  There is a direct dependence of $C_{\ell}$ on $V_{H}(r)$, which, in turn, has a direct dependence
on the mass density parameterization, $\delta(r)$, and on the void model used.

\section{Summary}
\label{secSummary}
In this paper, we investigated whether the underdensity in the local universe on scales extending to 
$\sim$200-300$~h^{-1}$~Mpc found by KBC is consistent with the latest SNe~Ia and kSZ data. 
The logic of our study was as follows:  (1) we used fixed CGBH void model mass density 
parameters to describe the KBC LLV, because there currently are insufficient $K$-band data to do 
a statistical analysis; 
(2) we found that the SNe Ia data are well described by our $\Lambda$LTB model with the adopted 
parameters and hence they do not rule out the existence of the KBC LLV;
(3) we found that the latest kSZ constraints from SPT and ACT observations are also compatible with our 
$\Lambda$LTB model parameterization.
We conclude that the significance of the $H_0$ discrepancy
with the CMB drops only modestly, from $3.4\sigma$ to $2.75\sigma$, due to the
presence of the KBC LLV. There is a corollary that a $\Lambda$LTB
model with more extreme parameters, e.g., a lower $\delta_{V}$, would not be compatible with
the SNe Ia data, because it would predict too large of a magnitude shift due to the void.

In detail, the steps of our analysis were as follows.  We chose a specific CGBH void model mass density 
parameterization of $\Delta r$ = 18.46~Mpc, $\delta_{V}$ = $-0.3$, and $r_{V}$ = 308~Mpc. 
We then studied the propagation of light through the KBC LLV by solving two ordinary differential 
equations in order to obtain the redshift dependence of the observables. 
We focused on two observables: the distance modulus and the linear kSZ effect.  First,
we tested the $\Lambda$LTB inhomogeneous cosmological model prediction for the luminosity-redshift relation inside 
the KBC LLV. We found this model description to be fully consistent with the 217 SNe~Ia of R16 lying in 
the redshift range $0.01 < z < 0.15$, which contains the void. Next, we assessed the implications of the KBC LLV in light 
of the $> 3 \sigma$ tension in measurements of the Hubble constant, $H_{0}$, by ``cosmic" and ``local" means.  
We presented evidence that a binning of the R16 SNe~Ia data inside ($0.0233 <z< 0.07$) and outside 
($0.07 <z< 0.15$) of the KBC LLV shows a value of $H_{0}$ that is 1.27$\pm$0.59~km~{$\rm s^{-1}~\rm Mpc^{-1}$} 
higher inside than outside.
This reduces the discrepancy with the cosmic $H_0$ from \textit{Planck} from $3.4\sigma$ to $2.75\sigma$.

Second, we examined the constraints from the linear kSZ-induced temperature fluctuations.
Previous work invoked the linear kSZ constraints from ACT and SPT 
to constrain the parameter space of GLV models with a 95\% confidence upper limit on the CMB secondary anisotropy power.  
The Doppler anisotropy term depends explicitly on the differing expansion rates via the LTB longitudinal Hubble parameter 
$H(t, r)$, inside and outside of the KBC LLV.  
We simplified the formula for the kSZ-induced temperature fluctuations by invoking the Limber approximation.  
Then, using the latest more stringent SPT (2014, 2015) constraints,
we generated linear kSZ constraining contours for the parameter space of CGBH void models.  
After comparing the CGBH void model parameters for the KBC LLV with 
these linear kSZ constraint contours, we concluded that the linear kSZ constraints imposed on the 
earlier set of GLV models, in addition to the more recent constraints from the SPT, are fully compatible with the 
existence of the KBC LLV.

\acknowledgements
We thank the anonymous referee for a helpful report that improved the manuscript.
We thank Adam Riess for his insightful comments on this study and for generously providing us with his SNe~Ia dataset that made our study feasible.  We thank Dan Scolnic for helpful information about the SuperCal SNe~Ia data.  B. L. H. was supported by the National Space Grant College and Fellowship Program and the Wisconsin Space Grant Consortium under NASA Training Grant  $\#$NNX15AJ12H, and by the UW-Madison Sophomore Research Fellowship under grants from the Brittingham Fund and the Kemper K. Knapp Bequest with additional support from the UW System and the UW-Madison Provost's Office.  A. J. B. gratefully acknowledges support from the John Simon Guggenheim Memorial Foundation and the Trustees of the William F. Vilas Estate.



\end{document}